# COVID-19 Detection Using CT Image Based On YOLOv5 Network

Ruyi Qu[1],
University of Toronto,
Toronto, Canada,
ruyi.qu@mail.utoronto.ca,

Yi yang[1],
University of Edinburgh,
Edinburgh, United of Kingdom
1902138171@qq.com,

Yuwei Wang[2],
Xidian University,
Xi'an, China,
951224105@qq.com

The COVID-19 pandemic has broken down the global medical order tremendously, we urgently need an efficient treatment. Computer aided diagnosis (CAD) increases diagnosis efficiency, helping doctors providing a quick and confident diagnosis, it has played an important role in the treatment of COVID-19. In our task, we solve the problem about abnormality detection and classification. The dataset provided by Kaggle platform and we choose YOLOv5 as our model. We introduce some methods on objective detection in the related work section, the objection detection can be divided into two streams: one- stage and two stage. The representational model are Faster RCNN and YOLO series. Then in the section III we describe YOLOv5 model in the detail. Compared Experiment and results are shown in section IV. We choose mean average precision (mAP) as our experiments' metrics, and the higher (mean )mAP is, the better result the model will gain. mAP@0.5 of our YOLOv5s is 0.623 which is 0.157 and 0.101 higher than Faster RCNN and EfficientDet respectively.

## I. Introduction

The COVID-19 pandemic has broken down the global medical order tremendously, which triggers an urgent need to improve the level of medical care. [1] Normally, COVID-19 diagnosis can be diagnosed via polymerase chain reaction, costing a long time almost 12 hours. If we need to get the quick result, the chest radiographs, obtained in minutes, can be an optimal choice. [2]
Because COVID-19 looks similar to other viral and bacterial pneumonias on chest radiographs, which is difficult to diagnose. There are some medical misdiagnosis when radiologists make the diagnosis. Computer aided diagnosis (CAD)[3] increases diagnosis efficiency, helping doctors providing a quick and confident diagnosis, it has played an important role in the treatment of COVID-19.
The computer vision technology mainly effects in the domain related with images, such as CT, MRI and PET scans. In this paper, we focus on the radiographs categorization, which contains three kinds: the negative for pneumonia, typical, indeterminate and atypical COVID-19. We describe the related objection detection methods in section II. And section III shows our model for this task: YOLO v5 while section IV introduces the experiment results and compared experiments. In the last section V, we draw the conclusion and proposed the further adjustment in the future.

## II. Related Work

Numerous approaches on the objection detection have been proposed in recent years, and there are two mainly traditional streams. The mainstream target detection algorithm based on deep learning is divided into two-stage target detection algorithm and single-stage target detection algorithm according to the generation stage of candidate box.[4][5]

*1) Two-stage method*

In view of the low precision of traditional target detection algorithms, Girshick et al. [6] proposed R-CNN, a goal detection algorithm based on deep learning, which pioneered a precedent. Ren et al. [7] proposed the Fast R-CNN algorithm, on the basis of Fast R-CNN added the candidate window network (region proposal network, RPN), the candidate window network by setting different scale anchors (anchor) to extract candidate boxes, instead of selective search and other traditional candidate box generation methods, to achieve end-to-end network training, improve the network computing speed. The Faster R-CNN network consists of convolution layers, RPN networks, ROI of interest, classification, and regression layers

*2) One-stage method*

In view of the inefficiency of the two-stage target detection algorithm, YOLO (you only look once) v1 [8]discarded the candidate box extraction branch in the algorithm and directly implemented feature extraction, candidate box classification and regression in the same branchless deep convolution network, making the network structure simple and increasing the detection speed from 7 frames/s of Fast R-CNN to 45 frames/s. So that the goal detection algorithm based on deep learning can meet the needs of real-time detection tasks under the computing power of the time. With the emergence of YOLov1, the goal detection algorithm based on deep learning begins to have two and single stages.
- Our Contribution
  ✓ We applied the model YOLOv5 for objective detection.
  ✓ We introduce our dataset and do some analysis.
  ✓ In the experiments process, we do the comparing experiments and the result shows that our model performed better than the other models.

### III. METHODOLOGY

The whole structure of YOLOv5 is shown in the following figure 1:

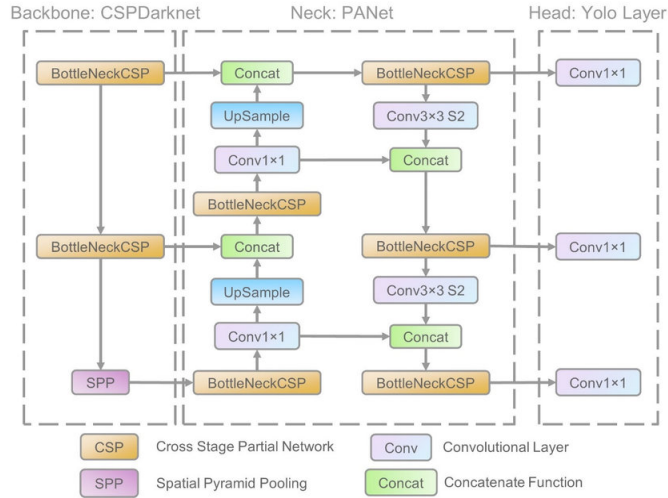

Figure 1: YOLOv5 Architecture

Backbone structure is the network core of YOLOv5 technology, which aims to extract information for utilization and processing in the process of image input. In the network structure model, gradient information often has many repetitive problems, which can be solved by combining CSPNet technology, so that gradient changes and feature diagrams can be integrated together, and the number of parameters of the model and the FLOPS value of the model are controlled in a lower range, thus improving the accuracy and efficiency of reasoning, and also achieving the goal of reducing the size of the model.

PANet is based on the Musk R-CNN network framework and the FPN network framework, and the dissemination of information is optimized and improved. In the process of extracting features, the network enhances the bottom-up path and improves the propagation of the underlying features. For the third network path, the data information characteristic mapping in the previous network stage is the input of this stage, operated according to the 3×3 convolution method, and the characteristic mapping value of each stage output will be directly connected with the path information of the same stage, effectively enhancing the joint utilization of high-level and low-level information. The damaged information paths between regions and features can also be recovered quickly under the action of adaptive feature pools, and the candidate regions are aggregated at each feature layer to avoid random allocation.

The basic components of YOLOv5 are five parts:

✓ The CBL-CBL module consists of the Conv-BN-Leaky_relu activation function.
✓ Res unit-drawing on the residual structure of the ResNet network is used to build a deep network, and CBM is a submodule in the residual module.
✓ CSP1_X: drawing on the CSPNet network structure, the module consists of CBL module, Res unint module, convolution layer, Concate.
✓ CSP2_X: Drawing on the CSPNet network structure, the module consists of a convolution layer and X Res unint module Concate.
✓ The Focus-Focus structure first puts multiple slide results into concat and then feeds them into the CBL module.
✓ SPP: multi-scale feature fusion is carried out using the largest pooling methods of 1×1, 5×5, 9×9 and 13×13

### IV. METHODOLOGY

- **Experimental Data**

Our task is to identify and localize COVID-19 abnormalities on chest radiograph. The train dataset comprises 6364 chest scans in DICOM format. All images were labeled by a panel of experience radiologists. We have five columns in the train_study_level.

| Id | Unique study identifier |
|---|---|
| Negative for pneumonia | Study is negative: 1 Otherwise: 0 |
| Typical appearance | Study has this appearance: 1; Otherwise: 0 |
| Indeterminate appearance | Study has this appearance: 1; Otherwise: 0 |
| Atypical appearance | Study has this appearance: 1; Otherwise: 0 |
| HEIGHT | 512 |
| WIDTH | 512 |
| CHANNELS | 3 |
| BATCH_SIZE | 32 |
| EPOCHS | 25 |

- **Experimental setting**

From the confusion matrix, we can get the how to calculate the precision and recall. We can draw the PR curve and compute the area under the curve to judge the model's performance.
In the objection detection task, we focus on the intersection over union (IOU).

$$IOU = \frac{area(B\_p \cap B\_gt)}{area(B\_p \cup B\_gt)}$$

B_gt represents the actual ground frame (Ground Truth, GT) of the target, and B_p represents the predicted frame. By calculating the IOU of the two, it can be judged whether the predicted detection frame meets the conditions.

From the basic definition, we can calculate the mAP. mAP is the average of AP. By adjusting different confidence thresholds, we can get different predictions, also means that we have different accuracy

- **Experimental Results and Analyses**

An example of input figures and output figures are shown as follows:

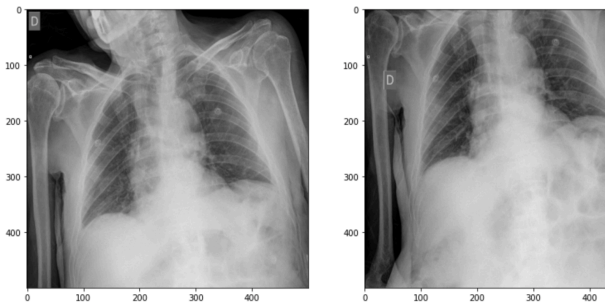

Figure2: Input Figure

- **Metrics**

In our task, the important metrics we focus is the (mean average precision) mAP score.[9]

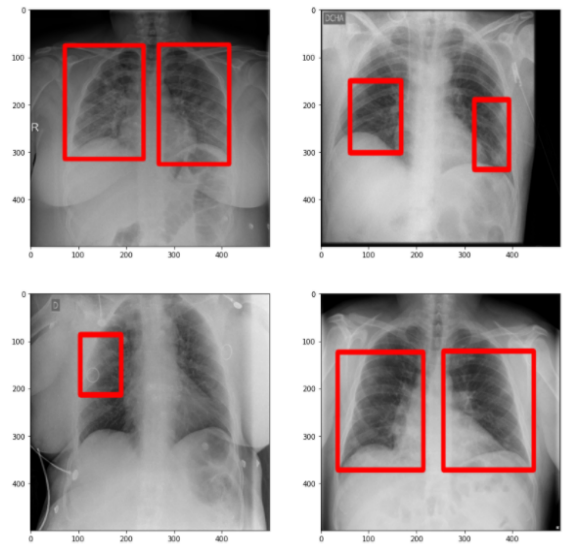

Figure 3: Output Figure

And the body part examined distribution are shown in the following figure 5, we can see that the chest is the highest body distribution examined is chest.
To compare our model's performance, we choose serval models do the compared experiments, such as Faster-RCNN and EfficientDet. We using the same metric mAP@0.5 And same datasets. The result is shown in table 1.

| Models | mAP@0.5 |
|---|---|
| Yolo v5s | 0.623 |
| Faster RCNN | 0.466 |
| EfficientDet | 0.522 |

Table 1: compared experiments

Yolov5s networks are the least deep and the smallest width of feature maps in the Yolov5 series. The latter three are on this basis to deepen, and constantly widen. EfficientDet[10]proposed to model scale the network such as the backbone of the detector. We know that the higher mAP is, the better result the model will gain. mAP@0.5 of our YOLOv5s is 0.623 which is 0.157 and 0.101 higher than Faster RCNN and EfficientDet respectively.

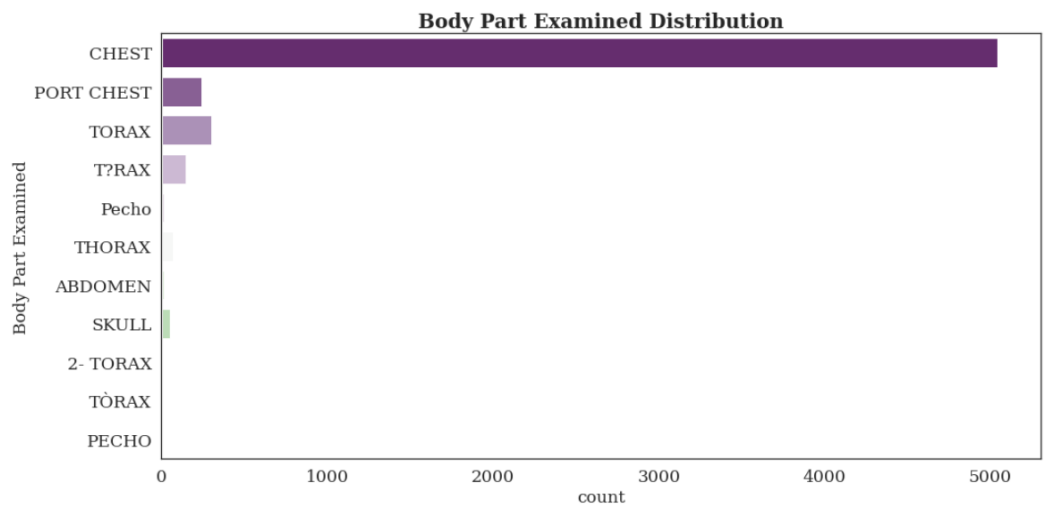
Figure 4: body part examined distribution

## V. CONCLUSION

Above, we have solved the task on abnormality detection, and this section will conclude our work. In our task, we solve the problem about abnormality detection and classification. The dataset provided by Kaggle platform and we choose YOLOv5 as our model. We introduce some methods on objective detection in the related work section, the objection detection Then in the section III we describe YOLOv5 model in the detail. Compared Experiment and results are shown in section IV. We choose mean average precision (mAP) as our experiments' metrics, and the higher (mean )mAP is, the better result the model will gain. mAP@0.5 of our YOLOv5s is 0.623 which is 0.157 and 0.101 higher than Faster RCNN and EfficientDet respectively.